\documentclass[10pt,aps,amsmath,amssymb, pre, twocolumn, superscriptaddress,hidelinks,showkeys]{revtex4-1}
\usepackage[utf8]{inputenc}
\usepackage[utf8]{inputenc}
\usepackage{wrapfig,lipsum,booktabs,amsmath,tikz,color,graphicx, float, pgfplots, subfigure, physics}
\usepackage[pdfpagelabels]{hyperref}
\usepackage{mathtools}
\usepackage{epstopdf}
\usepackage{silence}
\usepackage{upgreek}
\usepackage{enumerate}
\usepackage{enumitem}
\usepackage{breqn}
\pgfplotsset{compat=1.14}
\usepackage[activate={true,nocompatibility},final,tracking=true,kerning=true,factor=1100,stretch=10,shrink=10]{microtype}
\usepackage{etoolbox}

\usepackage{bm}
\usepackage{IEEEtrantools}
\makeatletter
\patchcmd\frontmatter@PACS@format{\addvspace{11\p@}}{}{}{}
\pretocmd\frontmatter@keys@format{\addvspace{11\p@}}{}{}
\patchcmd{\titleblock@produce}
  {\@pacs@produce\@pacs\@keywords@produce\@keywords}
  {\@keywords@produce\@keywords\@pacs@produce\@pacs}
  {}{}
\preto\maketitle{%
  \begingroup\lccode`~=`,
  \lowercase{\endgroup
  \let\saved@breqn@active@comma~
  \let~}\active@comma 
}
\appto\maketitle{%
  \begingroup\lccode`~=`,
  \lowercase{\endgroup
  \let~}\saved@breqn@active@comma 
}
\makeatother

\definecolor{applegreen}{rgb}{0, 0.5, 0.}

\begin{document}
\title{Diffuse interface models of solidification with convection: The choice of a finite interface thickness}
\author{Amol Subhedar}
\affiliation{Institute for Digital Materials Science, Karlsruhe University of Applied Sciences, Moltkestraße 30,  76133 Karlsruhe, Germany} 
\author{Peter K. Galenko}
\affiliation{Physikalisch-Astronomische Fakult{\"a}t, Friedrich-Schiller-Universit{\"a}t-Jena,  07743 Jena, Germany} 
\affiliation{Department of Theoretical and Mathematical Physics, Laboratory of Multi-Scale Mathematical Modeling, Ural Federal University, Lenin ave., 51, Ekaterinburg, 620000, Russian Federation }
\author{Fathollah Varnik}
\email{fathollah.varnik@rub.de}
\affiliation{Interdisciplinary Centre for Advanced Materials Simulation, Ruhr-Universit{\"a}t Bochum, Universit{\"a}tsstrasse 150, 44780 Bochum, Germany}

\begin{abstract}
 The thin interface limit aims at minimizing the effects arising from a numerical interface thickness, inherent in diffuse interface models of solidification and microstructure evolution such as the phase field model.
While the original formulation of this problem is restricted to transport by diffusion, we consider here the case of melt convection. {\textcolor{black}{Using}} an analysis of the coupled phase field-fluid dynamic equations, we show here that
 such a thin interface limit does also exist if transport contains both diffusion and convection. This prediction is tested by comparing simulation studies, which make use of the thin-interface condition, with an analytic sharp-interface theory for
dendritic tip growth under convection.
\end{abstract}


\date{\today}
\maketitle

\section{Introduction}
Letting aside critical phenomena, physical interfaces often have a width in the nanometer range. For problems on the mesoscale (i.e., dealing with micrometers or larger scales),
this thickness is negligible and the physical interface can safely be approximated to as a mathematically sharp boundary separating the phases of interest. The major aim of modeling at the mesoscale
 is thus to solve problems involving a sharp interface (SI). On the other hand, in the past thirty years, the so-called diffuse interface models such as the phase field (PF) approach~\cite{Boettinger2002,Steinbach2009} have proved quite
 powerful in studying solidification and microstructure evolution. These models involve a finite interface thickness, $W$, which, in view of the above mentioned fact, is of a numerical nature.
Ideally, one would like to minimize the effect of this numerical parameter.
{\textcolor{black}{The}} equivalence of a PF model of solidification to the SI formulation {\textcolor{black}{was established by Caginalp~\cite{Caginalp1989}}}
as the diffuse interface becomes progressively narrow ($W\to 0$). This ideal limit, however, is numerically quite expensive and often impractical. A major advancement was achieved in the mid 1990s
 by Karma and Rappel~\cite{Karma1996} with the so-called thin-interface limit for problems involving diffusive transport. They found that instead of vanishing interface thickness,
 it is sufficient to have $W$ small compared to the diffusion length of the solidification problem. \textcolor{black}{This diffusion length is defined as {\textcolor{black}{$L_{\text{d}} = \frac{D}{V}$}}, where $D$ is the thermal 
 diffusivity  and $V$ is the normal velocity of the interface. }

\section{Thin interface analysis in the presence of flow}

To account for transport due to convection in the solidification phenomena, a couple of melt flow and PF couplings have been proposed and analyzed. Anderson \textit{et al.}
performed~\cite{Anderson2001} a sharp interface asymptotics of a PF model where the viscosity of liquid melt-solid interface diverges while approaching solid end of the interface
(known as the variable viscosity model~\cite{Tonhardt1998,Britta2000,Anderson2000}). Beckermann \textit{et al.}~\cite{Beckermann1999} proposed a dissipative drag force ansatz that acts a momentum sink within the liquid melt-solid interface.
The strength of such a dissipative force, that is suitable for wide ranges of interface width to characteristics flow length ratio to ensure no-slip boundary condition, is then termed as an optimum coupling parameter $h^*$. Due to numerical simplicity of this approach,
many researchers have employed it for the simulation studies~\cite{Tong2001A,Jeong2001,Medvedev2006}, notwithstanding the fact that, a formal thin interface limit, for both of these coupling, has not been established.
The goal of present work is to summarize our findings on the existence of a  thin interface limit in such a case.

To keep the analysis tractable, \textcolor{black}{anisotropies} of the surface energy and kinetic coefficient are neglected. Diffusion coefficients and densities of the liquid and solid phases are assumed to be identical. Due to this equal density assumption, the melt velocity in a direction normal to the interface vanishes, thus simplifying the analysis. The growing solid is assumed to be stationary and does not move under the forces exerted by melt flow. Special attention is paid to ensure the no-slip boundary condition.

We introduce the following notation: $u$ is the reduced temperature field $u = \frac{T -T_{\text{m}}}{L/C_{\text{p}}}$, $T$ is temperature, $T_{\text{m}}$ is melting temperature, 
$L/C_{\text{p}}$ is the so-called hypercooling limit, $\delta$ is capillary length, $\beta$ is  kinetic coefficient,  $\mathbf{w}$ is the melt velocity, $\rho$ is density, $\mu_{\text{l}}$ is dynamic viscosity of the melt, $p$ is pressure, $\mathbf{g}$ is acceleration due to gravity. The phase field {\textcolor{black}{($\varphi$)}} and reduced temperature field equations are,
 \begin{eqnarray}
\tau \frac{\partial \varphi}{\partial t} = W^2\nabla^2 \varphi - f'(\varphi) - {\textcolor{black}{A_1}}\frac{W}{\delta} u g'(\varphi),\label{Eq:anPhi}\\
 \frac{\partial u}{\partial t} + \mathbf{w}\cdot \nabla u = D\nabla ^2 u + \frac{1}{2}\frac{\partial \varphi}{\partial t}, \label{Eq:anu}
\end{eqnarray}
where  $f'(\varphi) = -\varphi + \varphi^3$ is the well-known double well potential corresponding to phase field with values $\varphi=-1,+1$ in the liquid and solid phases, respectively. $g'(\varphi)=(1-\varphi^2)^2$ is an interpolating function that is non-zero only inside the interface,  ${\textcolor{black}{A_1}}$ is a numerical constant and $\tau$ is the relaxation time.  For the melt flow, we
first proceed with an improved version of the drag force model {\textcolor{black}{that ensures
Galilean invariance of the melt flow equations~\cite{Subhedar2015}}}. With this choice, the Navier-Stokes equations read,
\begin{equation}
\begin{split}
   \rho\left(\frac{\partial \mathbf{w}}{\partial t} + \mathbf{w}\cdot \nabla \mathbf{w}\right) &= -\nabla p + \mu_{\text{l}} \nabla ^2 \mathbf{w} + \rho(1-{\textcolor{black}{\gamma}} u) \mathbf{g}  \\
   &- h^{*}\mu_{\text{l}} \frac{H(\varphi)}{W^2}\mathbf{w}, \label{Eq:beckerFluid}
\end{split}
\end{equation}
where  ${\textcolor{black}{\gamma}}$ is a coefficient related to thermal expansion,
and $H(\varphi)$ is {\textcolor{black}{an}} interpolating polynomial with $H(\pm 1)=0$. \textcolor{black}{The description of melt flow dynamics is completed with the continuity equation,
\begin{equation}
 \frac{\partial \rho}{\partial t} +\nabla\cdot{\rho\mathbf{w}} = 0. \label{Eq:continuity}
\end{equation}}

The small parameter for the asymptotic expansion is identified as a ratio of interface width to  diffusion length, $\varepsilon = \frac{WV}{D} ={\textcolor{black}{\frac{W}{L_{\text{d}}}}}$. A curvilinear orthogonal system of coordinates that is attached to the moving interface, with unit vectors $\mathbf{\hat r}$ (normal to the interface) and $\mathbf{\hat s}$ (tangential)
is chosen to analyze the coupled set of equations. The scaled length  in a direction normal to the interface, $\frac{r}{\varepsilon}$, is denoted by $\eta$. The limit $\eta \to \pm\infty$ corresponds to the  liquid and solid side of the interface, respectively.

Melt flow is expanded for inner $\mathbf{w}$ (microscopic) and outer $\tilde{\mathbf{w}}$ (macroscopic)  variables, up to second order in $\varepsilon$ as
$\mathbf{w} \approx \mathbf{w}_0 + \varepsilon \mathbf{w}_1 + \varepsilon^2 \mathbf{w}_2 $ and
$ \tilde{\mathbf{w}} \approx \tilde {\mathbf{w}}_0 + \varepsilon\tilde {\mathbf{w}}_1 + \varepsilon^2 \tilde {\mathbf{w}}_2 $. {\textcolor{black}{A similar expansion is used for $\varphi$ and $u$, where $u_n, \varphi_n$ denotes
order of approximation in $\varepsilon$ for integer $n$.}}
The macroscopic melt velocity can be Taylor expanded in the direction normal to the interface around the position of a hypothetical sharp interface at $r = 0$ as follows~\cite{Subhedar2019},
\begin{equation}
\begin{split}
 \tilde{\mathbf{w}} &= \tilde {\mathbf{w}}_0(0) + \varepsilon\left(\eta\frac{\partial \tilde {\mathbf{w}}_0(0)}{\partial r}+\tilde {\mathbf{w}}_1(0) \right) \\
 &+\varepsilon^2\left(\frac{\eta^2}{2}\frac{\partial^2 \tilde {\mathbf{w}}_0(0)}{\partial r^2} + \eta\frac{\partial \tilde {\mathbf{w}}_1(0)}{\partial r}+\tilde {\mathbf{w}}_2(0) \right)\textcolor{black}{.} \label{Eq:taylorU}
\end{split}
\end{equation}

In the present case, the no-slip boundary condition at the liquid-solid interface can be written as $\mathbf{\tilde{w}}^{-}{(0)} = 0$.
The superscript $-$ denotes the quantity evaluated at the interface when approached from the solid side of the phase.
${\mathbf{\tilde w}}(0)$ reminds us that these macroscopic quantities are evaluated at the sharp interface position, $r=0$, which coincides with the center of the diffuse interface. From the no-slip boundary condition we conclude that
$\frac{\partial^k \tilde {\mathbf{w}}^{-}_0(0)}{\partial r^k}  = 0$, for positive natural integer $k$.
We  denote the normal and tangential components of the melt velocity by ${w}^s$ and ${w}^r$. 
 {\textcolor{black} {We write the continuity and momentum balance 
equations for the melt flow dynamics as,}}
\textcolor{black}{\begin{equation}
  \frac{1}{\varepsilon}\partial_{\eta}{w}^r + \kappa {w}^r +\frac{1}{1+\varepsilon \eta\kappa}\partial_{s}{w}^s  = 0,\label{Eq:conScaledCoor} 
\end{equation}}

\begin{equation}
\begin{split}
 &\rho\left(\frac{{w}^r}{\varepsilon}\partial_{\eta}  + \frac{{w}^s}{1+\varepsilon \eta\kappa} \partial_s\right)w^r - \rho\frac{{w^s}^2\kappa}{1+\eta\varepsilon \kappa} = - \frac{1}{\varepsilon}\partial_{\eta} p  \\ 
 &  + \mu_{\text{l}} \left(\frac{1}{\varepsilon^2}\partial_{\eta\eta} + \frac{1}{\varepsilon} \kappa \partial_{\eta} \right)w^r + \rho(1-\gamma u) g^r - h^{*}\mu_{\text{l}} \frac{H(\varphi)}{L_{\text{d}}^2\varepsilon^2}{w^r}, 
\end{split}\label{Eq:movU}
\end{equation}

\begin{equation}
\begin{split}
 &\rho\left(\frac{{w}^r}{\varepsilon}\partial_{\eta}  + \frac{{w}^s}{1+\varepsilon \eta\kappa} \partial_s\right)w^s - \rho\frac{{w^s}w^r\kappa}{1+\eta\varepsilon \kappa}  =  - \frac{1}{1+\varepsilon\eta\kappa}\partial_{s} p  \\
 &+ \mu_{\text{l}}\frac{1}{\varepsilon^2} \left(\partial_{\eta\eta} + {\varepsilon} \kappa \partial_{\eta} \right)w^s  + \rho(1-{\textcolor{black}{\gamma}}  u) g^s  - h^{*}\mu_{\text{l}} \frac{H(\varphi)}{L_{\text{d}}^2\varepsilon^2}{w^s}, \\
\end{split}\label{Eq:movUT}
\end{equation}
{\textcolor{black}{where $\kappa$ is  interface curvature and $g^r,g^s$ are normal and tangential components of the gravity.}}
\textcolor{black}{Noting that the densities of the melt and solid are the same, the variation of the normal component of the melt velocity $w^r$ across the interface is neglected. With this assumption, 
we proceed with analysis of Eq.~(\ref{Eq:conScaledCoor}) and Eq.~(\ref{Eq:movUT})}
 at successive orders of $\varepsilon$. At the second order, in combination with Eq.~(\ref{Eq:taylorU}), we obtain  $\dfrac{\partial^2 {\tilde{{w}}^{s-}_2}}{\partial r^2} =  -\dfrac{\rho(1-{\textcolor{black}{\gamma}}  u_0)g^s}{\mu_{\text{l}}}$. This means that the matching condition on the solid side of the interface is not satisfied in the second order of $\varepsilon$-expansion for the drag force model. In view of this result, we also examined
{\textcolor{black} {the}} variable viscosity model~\cite{Britta2000}.
For this choice, through a similar analysis, we obtain $\frac{\partial \tilde {\mathbf{w}}_1^{-}(0)}{\partial r}  = 0 , \frac{\partial^2 \tilde {\mathbf{w}}_0^{-}(0)}{\partial r^2} = 0 $. An extensive account of this approach
{\textcolor{black}{and  additional issues faced by the original drag force model~\cite{Beckermann1999} due to violation of the Galilean invariance}} can be found in Ref.~\cite{Subhedar2019}. 
We just here conclude that, the variable viscosity model can satisfy matching condition for inner and outer velocity fields, even in the presence of body forces like gravity. 
For both of these couplings the relation between phase field parameters $\tau=W^2\left( \frac{\beta}{\delta} + \frac{M}{D}\frac{W}{\delta}\right)$, that was originally devised for the diffusive transport~\cite{Karma1996}, remains valid.  
This relation is necessary to comply with macroscopic energy balance and Gibbs-Thomson relation at the interface. The constant M depends on the chosen forms of $f'(\varphi)$ and $g'(\varphi)$ in Eq.~(\ref{Eq:anPhi}).

\section{Numerical simulations}
To test the above analysis, we perform numerical simulations of a 2D dendrite, growing in the direction opposite to an externally imposed melt flow.
We include a fourfold surface energy anisotropy in \textcolor{black}{Eq.~(\ref{Eq:anPhi})} as described in Ref.~\cite{Karma1996}. 
\textcolor{black}{The phase field equation in this cases is,}
\begin{equation}
 \begin{split}  
 \tau \frac{\partial \varphi}{\partial t} &= \nabla \cdot (W^2 \nabla \varphi ) + \nabla \cdot \left(|\nabla \varphi|^2 W \frac{\partial W}{\partial \nabla \varphi}\right) + \varphi - \varphi^3 \\
 &-{\textcolor{black}{A_1}}\frac{W}{\delta} u (1-\varphi^2)^2, \label{Eq:pfAniso}
 \end{split}
\end{equation}
\textcolor{black}{where  $\tau = \tau_0a(\mathbf{n})^2,  W = W_0 a(\mathbf{n})$, $a(\mathbf{n})=  1 + \epsilon \cos(4 \theta)$ and $\theta$ is the
angle between normal to the interface and some fixed direction. $\tau_0$ and $W_0$ are the reference relaxation time and interface width. $\epsilon$ denotes the strength of surface energy anisotropy and a positive value
of $\epsilon$ is a necessary condition to achieve a steady state. This construction effectively ensures the capillary length $\delta$ to be of the form $\delta = \delta_0(1-15\epsilon\cos(4\theta))$.}
\textcolor{black}{In addition to Eq.~(\ref{Eq:pfAniso}), the heat transport and melt flow dynamics are solved with Eq.~(\ref{Eq:anu}),
Eq.~(\ref{Eq:beckerFluid}) and Eq.~(\ref{Eq:continuity})}. 

To compare the simulated growth velocity with the corresponding sharp-interface solution, we refer to a recently developed analytic {\textcolor{black}{Alexandrov-Galenko (AG)}} theory~\cite{Alexandrov2013}, which predicts,
\begin{equation}
\tilde V_{\text{g}}  = \frac{V_{\text{g}}\delta}{D} =   \frac{2\sigma_0 \epsilon^{7/4} P^2_{\text{g}}}{(1 + a_1 \sqrt{\epsilon} P_{\text{g}})^2}\left[1 + b\left(\frac{\alpha}{\epsilon^{3/4}}\right)^{\frac{11}{14}}\right]^{-1}.\label{Eq:velPredGalenko}
\end{equation}
Here: $b$, $a_1$, $\sigma_0$ are numerical constant\textcolor{black}{s}, 
$$\alpha = \frac{ a(\text{Re})  |\mathbf{w}_{\infty}|\delta}{4R V_{\text{g}}}, 
\quad a(\text{Re})= \sqrt{\frac{\text{Re}}{2\pi}}
\frac{\exp{(-\text{Re}/2)}}{\rm{erfc}(\sqrt{(\text{Re}/2})},$$ 
and $R$ is the tip radius of resulting steady state parabola ($|\mathbf{w}_{\infty}|$ is the far field melt velocity).
$V_{\text{g}}$ is the \textcolor{black}{steady state} growth velocity, \textcolor{black}{$\text{Re} = \frac{\rho R|\mathbf{w}_{\infty}|}{\mu_{\text{l}}}$ is the Reynolds number} and $P_{\text{g}}=V_{\text{g}}R/(2D)$ is the growth  P{\'e}clet number.

\begin{figure}
a)  \includegraphics[height=5cm]{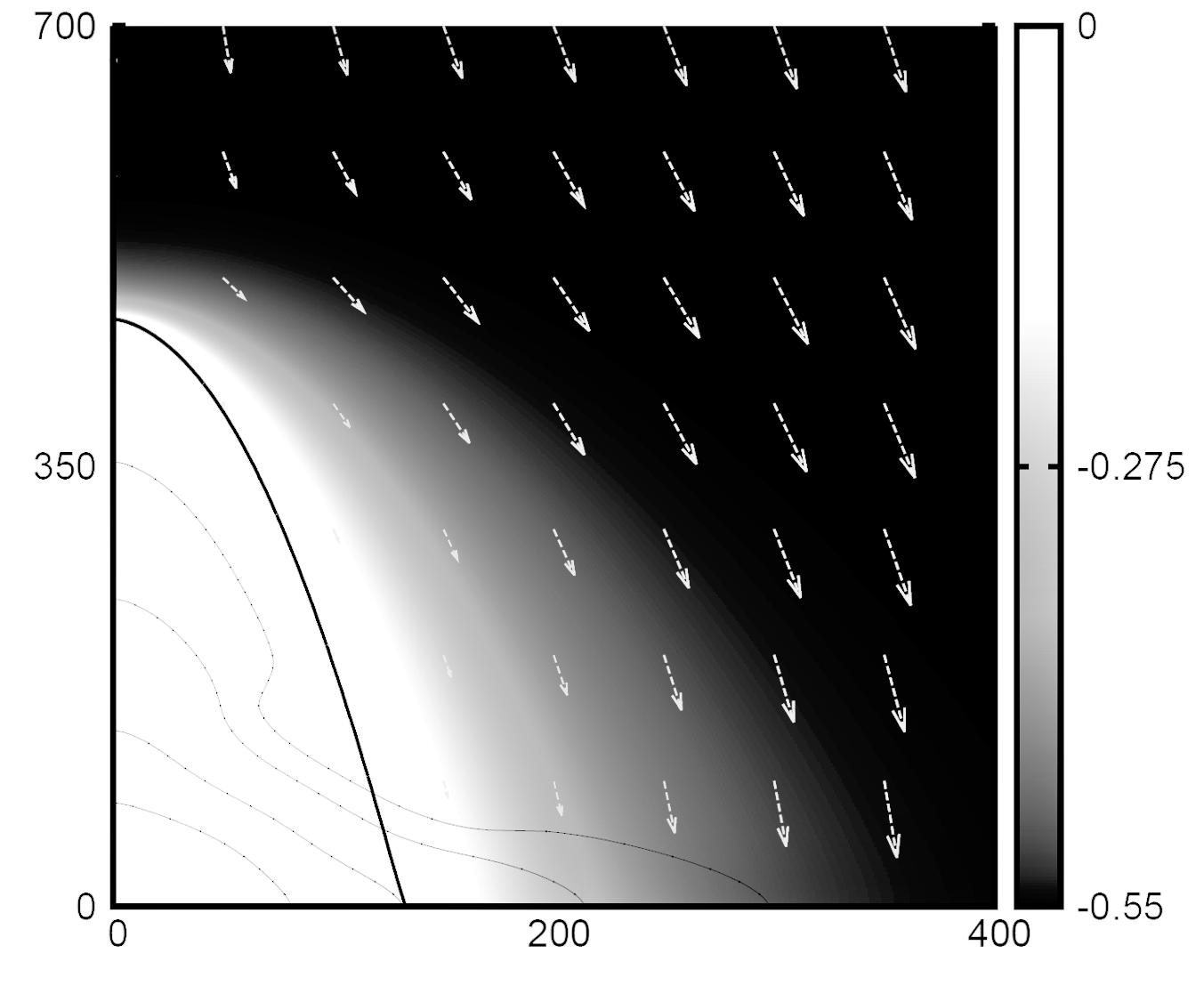}\hspace*{10mm}\\
b)  \includegraphics[height=5.5cm]{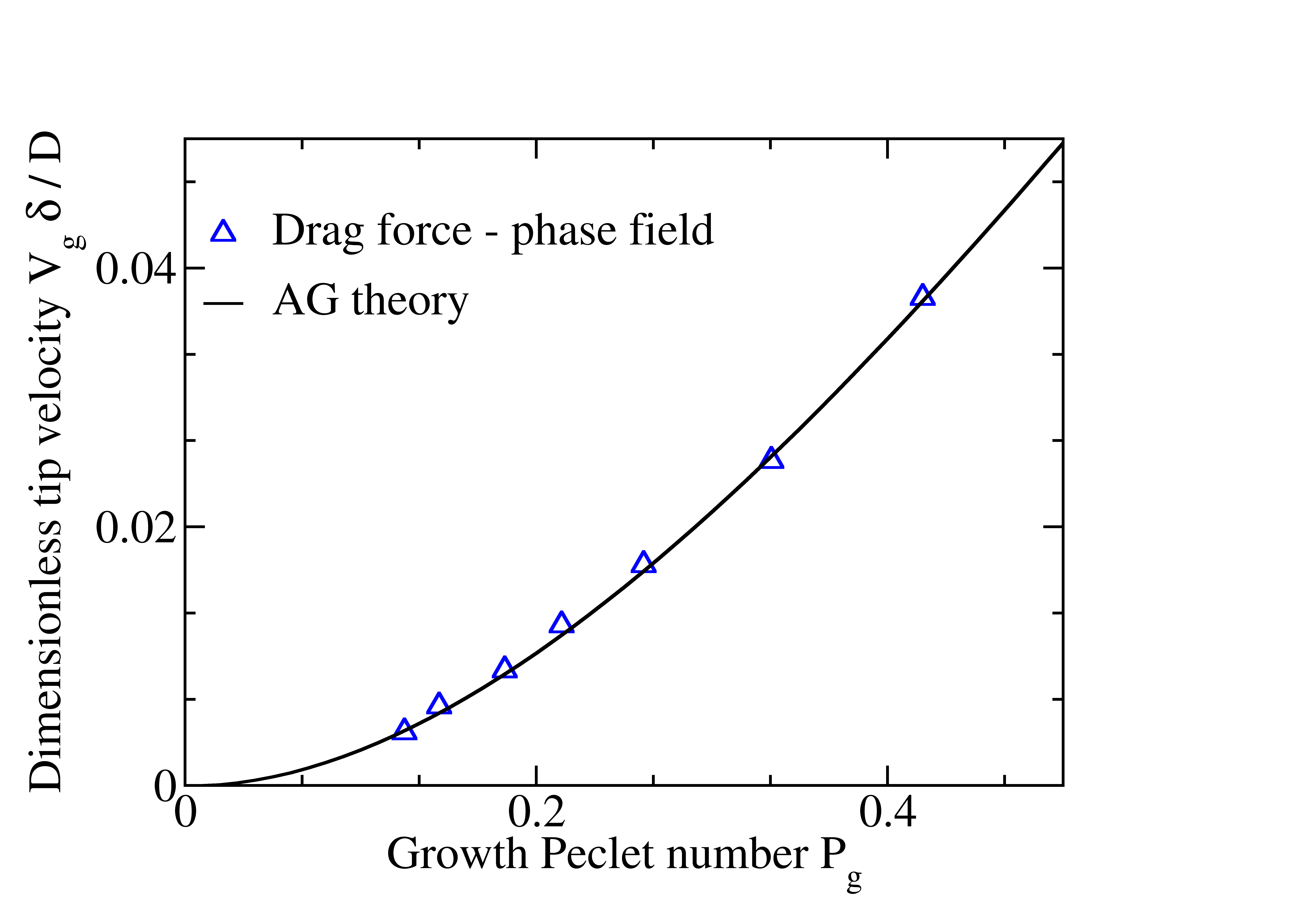}
\caption{a) A growing dendrite subject to convection. \textcolor{black}{For numerical solution, coupled system of equations Eq.~(\ref{Eq:anu}), Eq.~(\ref{Eq:beckerFluid}), Eq.~(\ref{Eq:continuity}) and  Eq.~(\ref{Eq:pfAniso})
are solved.} {\textcolor{black}{The arrow head shows direction while  length shows strength of melt flow velocity}}. b) The dimensionless tip velocity versus the growth  P{\'e}clet number $P_{\text{g}}$ {\textcolor{black}{calculated by the present PF model (triangles) and AG theory~\cite{Alexandrov2013}}}.
}
\label{Fig:theoryCompare}
\end{figure}

{\textcolor{black}{Figure~(1a) shows a typical growth of a dendrite along with iso-temperature curves and velocity vector arrows surrounding the dendrite. }}
Figure~(\ref{Fig:theoryCompare}{\textcolor{black}{b}}) compares  the scaled velocity $\tilde V_{\text{g}}$ versus $P_{\text{g}}$ for the PF simulations and the AG theory, showing excellent agreement between the two. Simulations with higher flow velocities confirm this agreement further (results not shown).
\textcolor{black}{This agreement suggests that neglecting anisotropic terms in thin interface asymptotics does not alter the main conclusion regarding the independence of the simulation results on the interface thickness.}

\section{Conclusions}
A thin-interface analysis of the phase field equations in the presence of melt convection is provided. It is shown that, as in the case of diffusive transport, the thickness of the diffuse interface can be chosen such that its effects on the obtained results are minimized. This prediction is verified by a comparison of the numerical simulation results for dendritic tip velocity and an analytic theory, which accounts for flow effects.
As an outlook for further work, it shall be noted that, unlike the temperature or the solute fields, the melt velocity identically vanishes in the solid domain.
{\textcolor{black}{The  non-vanishing normal gradients of the tangential velocity contribute to the shear stress tensor that generates an equal and opposite resulting force on the growing solid. The solid structure,
when allowed to move, in turn\textcolor{black}{,} influences the melt flow field and thereby transport of heat and solute. In such a case, the unbalanced shear stresses on solid might play an important role.}}

\section*{Acknowledgments}
P.\,K.\,G. acknowledges the support by the European Space Agency (ESA) under research project MULTIPHAS (AO-2004) and the German Aerospace Center (DLR) Space Management under contract No. 50WM1541 and also from the Russian Science Foundation under the project no. 16-11-10095. A.~S. and F.~V. acknowledges financial support by the German Research Foundation (DFG) under the project number Va205/17-1.

\bibliography{convectionThinInterfaceEPJST}{}

\begin{thebibliography}{15}%
\makeatletter
\providecommand \@ifxundefined [1]{%
 \@ifx{#1\undefined}
}%
\providecommand \@ifnum [1]{%
 \ifnum #1\expandafter \@firstoftwo
 \else \expandafter \@secondoftwo
 \fi
}%
\providecommand \@ifx [1]{%
 \ifx #1\expandafter \@firstoftwo
 \else \expandafter \@secondoftwo
 \fi
}%
\providecommand \natexlab [1]{#1}%
\providecommand \enquote  [1]{``#1''}%
\providecommand \bibnamefont  [1]{#1}%
\providecommand \bibfnamefont [1]{#1}%
\providecommand \citenamefont [1]{#1}%
\providecommand \href@noop [0]{\@secondoftwo}%
\providecommand \href [0]{\begingroup \@sanitize@url \@href}%
\providecommand \@href[1]{\@@startlink{#1}\@@href}%
\providecommand \@@href[1]{\endgroup#1\@@endlink}%
\providecommand \@sanitize@url [0]{\catcode `\\12\catcode `\$12\catcode
  `\&12\catcode `\#12\catcode `\^12\catcode `\_12\catcode `\%12\relax}%
\providecommand \@@startlink[1]{}%
\providecommand \@@endlink[0]{}%
\providecommand \url  [0]{\begingroup\@sanitize@url \@url }%
\providecommand \@url [1]{\endgroup\@href {#1}{\urlprefix }}%
\providecommand \urlprefix  [0]{URL }%
\providecommand \Eprint [0]{\href }%
\providecommand \doibase [0]{http://dx.doi.org/}%
\providecommand \selectlanguage [0]{\@gobble}%
\providecommand \bibinfo  [0]{\@secondoftwo}%
\providecommand \bibfield  [0]{\@secondoftwo}%
\providecommand \translation [1]{[#1]}%
\providecommand \BibitemOpen [0]{}%
\providecommand \bibitemStop [0]{}%
\providecommand \bibitemNoStop [0]{.\EOS\space}%
\providecommand \EOS [0]{\spacefactor3000\relax}%
\providecommand \BibitemShut  [1]{\csname bibitem#1\endcsname}%
\let\auto@bib@innerbib\@empty
\bibitem [{\citenamefont {Boettinger}\ \emph {et~al.}(2002)\citenamefont
  {Boettinger}, \citenamefont {Warren}, \citenamefont {Beckermann},\ and\
  \citenamefont {Karma}}]{Boettinger2002}%
  \BibitemOpen
  \bibfield  {author} {\bibinfo {author} {\bibfnamefont {W.~J.}\ \bibnamefont
  {Boettinger}}, \bibinfo {author} {\bibfnamefont {J.~A.}\ \bibnamefont
  {Warren}}, \bibinfo {author} {\bibfnamefont {C.}~\bibnamefont {Beckermann}},
  \ and\ \bibinfo {author} {\bibfnamefont {A.}~\bibnamefont {Karma}},\ }\href
  {\doibase 10.1146/annurev.matsci.32.101901.155803} {\bibfield  {journal}
  {\bibinfo  {journal} {Annual Review of Materials Research}\ }\textbf
  {\bibinfo {volume} {32}},\ \bibinfo {pages} {163} (\bibinfo {year}
  {2002})}\BibitemShut {NoStop}%
\bibitem [{\citenamefont {Steinbach}(2009)}]{Steinbach2009}%
  \BibitemOpen
  \bibfield  {author} {\bibinfo {author} {\bibfnamefont {I.}~\bibnamefont
  {Steinbach}},\ }\href {\doibase
  https://doi.org/10.1016/j.actamat.2009.02.004} {\bibfield  {journal}
  {\bibinfo  {journal} {Acta Materialia}\ }\textbf {\bibinfo {volume} {57}},\
  \bibinfo {pages} {2640 } (\bibinfo {year} {2009})}\BibitemShut {NoStop}%
\bibitem [{\citenamefont {Caginalp}(1989)}]{Caginalp1989}%
  \BibitemOpen
  \bibfield  {author} {\bibinfo {author} {\bibfnamefont {G.}~\bibnamefont
  {Caginalp}},\ }\href {\doibase 10.1103/PhysRevA.39.5887} {\bibfield
  {journal} {\bibinfo  {journal} {Phys. Rev. A}\ }\textbf {\bibinfo {volume}
  {39}},\ \bibinfo {pages} {5887} (\bibinfo {year} {1989})}\BibitemShut
  {NoStop}%
\bibitem [{\citenamefont {Karma}\ and\ \citenamefont
  {Rappel}(1996)}]{Karma1996}%
  \BibitemOpen
  \bibfield  {author} {\bibinfo {author} {\bibfnamefont {A.}~\bibnamefont
  {Karma}}\ and\ \bibinfo {author} {\bibfnamefont {W.-J.}\ \bibnamefont
  {Rappel}},\ }\href {\doibase 10.1103/PhysRevLett.77.4050} {\bibfield
  {journal} {\bibinfo  {journal} {Phys. Rev. Lett.}\ }\textbf {\bibinfo
  {volume} {77}},\ \bibinfo {pages} {4050} (\bibinfo {year}
  {1996})}\BibitemShut {NoStop}%
\bibitem [{\citenamefont {Anderson}\ \emph {et~al.}(2001)\citenamefont
  {Anderson}, \citenamefont {McFadden},\ and\ \citenamefont
  {Wheeler}}]{Anderson2001}%
  \BibitemOpen
  \bibfield  {author} {\bibinfo {author} {\bibfnamefont {D.}~\bibnamefont
  {Anderson}}, \bibinfo {author} {\bibfnamefont {G.}~\bibnamefont {McFadden}},
  \ and\ \bibinfo {author} {\bibfnamefont {A.}~\bibnamefont {Wheeler}},\ }\href
  {\doibase https://doi.org/10.1016/S0167-2789(01)00229-9} {\bibfield
  {journal} {\bibinfo  {journal} {Physica D: Nonlinear Phenomena}\ }\textbf
  {\bibinfo {volume} {151}},\ \bibinfo {pages} {305 } (\bibinfo {year}
  {2001})}\BibitemShut {NoStop}%
\bibitem [{\citenamefont {Tönhardt}\ and\ \citenamefont
  {Amberg}(1998)}]{Tonhardt1998}%
  \BibitemOpen
  \bibfield  {author} {\bibinfo {author} {\bibfnamefont {R.}~\bibnamefont
  {Tönhardt}}\ and\ \bibinfo {author} {\bibfnamefont {G.}~\bibnamefont
  {Amberg}},\ }\href {\doibase https://doi.org/10.1016/S0022-0248(98)00687-3}
  {\bibfield  {journal} {\bibinfo  {journal} {Journal of Crystal Growth}\
  }\textbf {\bibinfo {volume} {194}},\ \bibinfo {pages} {406 } (\bibinfo {year}
  {1998})}\BibitemShut {NoStop}%
\bibitem [{\citenamefont {Nestler}\ \emph {et~al.}(2000)\citenamefont
  {Nestler}, \citenamefont {Wheeler}, \citenamefont {Ratke},\ and\
  \citenamefont {Stöcker}}]{Britta2000}%
  \BibitemOpen
  \bibfield  {author} {\bibinfo {author} {\bibfnamefont {B.}~\bibnamefont
  {Nestler}}, \bibinfo {author} {\bibfnamefont {A.}~\bibnamefont {Wheeler}},
  \bibinfo {author} {\bibfnamefont {L.}~\bibnamefont {Ratke}}, \ and\ \bibinfo
  {author} {\bibfnamefont {C.}~\bibnamefont {Stöcker}},\ }\href {\doibase
  https://doi.org/10.1016/S0167-2789(00)00035-X} {\bibfield  {journal}
  {\bibinfo  {journal} {Physica D: Nonlinear Phenomena}\ }\textbf {\bibinfo
  {volume} {141}},\ \bibinfo {pages} {133 } (\bibinfo {year}
  {2000})}\BibitemShut {NoStop}%
\bibitem [{\citenamefont {Anderson}\ \emph {et~al.}(2000)\citenamefont
  {Anderson}, \citenamefont {McFadden},\ and\ \citenamefont
  {Wheeler}}]{Anderson2000}%
  \BibitemOpen
  \bibfield  {author} {\bibinfo {author} {\bibfnamefont {D.}~\bibnamefont
  {Anderson}}, \bibinfo {author} {\bibfnamefont {G.}~\bibnamefont {McFadden}},
  \ and\ \bibinfo {author} {\bibfnamefont {A.}~\bibnamefont {Wheeler}},\ }\href
  {\doibase https://doi.org/10.1016/S0167-2789(99)00109-8} {\bibfield
  {journal} {\bibinfo  {journal} {Physica D: Nonlinear Phenomena}\ }\textbf
  {\bibinfo {volume} {135}},\ \bibinfo {pages} {175 } (\bibinfo {year}
  {2000})}\BibitemShut {NoStop}%
\bibitem [{\citenamefont {Beckermann}\ \emph {et~al.}(1999)\citenamefont
  {Beckermann}, \citenamefont {Diepers}, \citenamefont {Steinbach},
  \citenamefont {Karma},\ and\ \citenamefont {Tong}}]{Beckermann1999}%
  \BibitemOpen
  \bibfield  {author} {\bibinfo {author} {\bibfnamefont {C.}~\bibnamefont
  {Beckermann}}, \bibinfo {author} {\bibfnamefont {H.-J.}\ \bibnamefont
  {Diepers}}, \bibinfo {author} {\bibfnamefont {I.}~\bibnamefont {Steinbach}},
  \bibinfo {author} {\bibfnamefont {A.}~\bibnamefont {Karma}}, \ and\ \bibinfo
  {author} {\bibfnamefont {X.}~\bibnamefont {Tong}},\ }\href {\doibase
  https://doi.org/10.1006/jcph.1999.6323} {\bibfield  {journal} {\bibinfo
  {journal} {Journal of Computational Physics}\ }\textbf {\bibinfo {volume}
  {154}},\ \bibinfo {pages} {468 } (\bibinfo {year} {1999})}\BibitemShut
  {NoStop}%
\bibitem [{\citenamefont {Tong}\ \emph {et~al.}(2001)\citenamefont {Tong},
  \citenamefont {Beckermann}, \citenamefont {Karma},\ and\ \citenamefont
  {Li}}]{Tong2001A}%
  \BibitemOpen
  \bibfield  {author} {\bibinfo {author} {\bibfnamefont {X.}~\bibnamefont
  {Tong}}, \bibinfo {author} {\bibfnamefont {C.}~\bibnamefont {Beckermann}},
  \bibinfo {author} {\bibfnamefont {A.}~\bibnamefont {Karma}}, \ and\ \bibinfo
  {author} {\bibfnamefont {Q.}~\bibnamefont {Li}},\ }\href {\doibase
  10.1103/PhysRevE.63.061601} {\bibfield  {journal} {\bibinfo  {journal} {Phys.
  Rev. E}\ }\textbf {\bibinfo {volume} {63}},\ \bibinfo {pages} {061601}
  (\bibinfo {year} {2001})}\BibitemShut {NoStop}%
\bibitem [{\citenamefont {Jeong}\ \emph {et~al.}(2001)\citenamefont {Jeong},
  \citenamefont {Goldenfeld},\ and\ \citenamefont {Dantzig}}]{Jeong2001}%
  \BibitemOpen
  \bibfield  {author} {\bibinfo {author} {\bibfnamefont {J.-H.}\ \bibnamefont
  {Jeong}}, \bibinfo {author} {\bibfnamefont {N.}~\bibnamefont {Goldenfeld}}, \
  and\ \bibinfo {author} {\bibfnamefont {J.~A.}\ \bibnamefont {Dantzig}},\
  }\href {\doibase 10.1103/PhysRevE.64.041602} {\bibfield  {journal} {\bibinfo
  {journal} {Phys. Rev. E}\ }\textbf {\bibinfo {volume} {64}},\ \bibinfo
  {pages} {041602} (\bibinfo {year} {2001})}\BibitemShut {NoStop}%
\bibitem [{\citenamefont {Medvedev}\ and\ \citenamefont
  {Kassner}(2005)}]{Medvedev2006}%
  \BibitemOpen
  \bibfield  {author} {\bibinfo {author} {\bibfnamefont {D.}~\bibnamefont
  {Medvedev}}\ and\ \bibinfo {author} {\bibfnamefont {K.}~\bibnamefont
  {Kassner}},\ }\href {\doibase 10.1103/PhysRevE.72.056703} {\bibfield
  {journal} {\bibinfo  {journal} {Phys. Rev. E}\ }\textbf {\bibinfo {volume}
  {72}},\ \bibinfo {pages} {056703} (\bibinfo {year} {2005})}\BibitemShut
  {NoStop}%
\bibitem [{\citenamefont {Subhedar}\ \emph {et~al.}(2015)\citenamefont
  {Subhedar}, \citenamefont {Steinbach},\ and\ \citenamefont
  {Varnik}}]{Subhedar2015}%
  \BibitemOpen
  \bibfield  {author} {\bibinfo {author} {\bibfnamefont {A.}~\bibnamefont
  {Subhedar}}, \bibinfo {author} {\bibfnamefont {I.}~\bibnamefont {Steinbach}},
  \ and\ \bibinfo {author} {\bibfnamefont {F.}~\bibnamefont {Varnik}},\ }\href
  {\doibase 10.1103/PhysRevE.92.023303} {\bibfield  {journal} {\bibinfo
  {journal} {Phys. Rev. E}\ }\textbf {\bibinfo {volume} {92}},\ \bibinfo
  {pages} {023303} (\bibinfo {year} {2015})}\BibitemShut {NoStop}%
\bibitem [{\citenamefont {Subhedar}\ \emph {et~al.}()\citenamefont {Subhedar},
  \citenamefont {Galenko},\ and\ \citenamefont {Varnik}}]{Subhedar2019}%
  \BibitemOpen
  \bibfield  {author} {\bibinfo {author} {\bibfnamefont {A.}~\bibnamefont
  {Subhedar}}, \bibinfo {author} {\bibfnamefont {P.}~\bibnamefont {Galenko}}, \
  and\ \bibinfo {author} {\bibfnamefont {F.}~\bibnamefont {Varnik}},\
  }\href@noop {} {\enquote {\bibinfo {title} {Thin interface limit of the
  double sided phase field model with convection},}\ }\bibinfo {note} {In
  preparation}\BibitemShut {NoStop}%
\bibitem [{\citenamefont {Alexandrov}\ and\ \citenamefont
  {Galenko}(2013)}]{Alexandrov2013}%
  \BibitemOpen
  \bibfield  {author} {\bibinfo {author} {\bibfnamefont {D.~V.}\ \bibnamefont
  {Alexandrov}}\ and\ \bibinfo {author} {\bibfnamefont {P.~K.}\ \bibnamefont
  {Galenko}},\ }\href {\doibase 10.1103/PhysRevE.87.062403} {\bibfield
  {journal} {\bibinfo  {journal} {Phys. Rev. E}\ }\textbf {\bibinfo {volume}
  {87}},\ \bibinfo {pages} {062403} (\bibinfo {year} {2013})}\BibitemShut
  {NoStop}%
\end{thebibliography}%
\bibliographystyle{apsrev4-1}
\end{document}